\documentclass[12pt]{article}

\usepackage{graphicx}
\usepackage{amsmath,amssymb,amsfonts}
\usepackage{hyperref}
\usepackage{lmodern}
\usepackage{amssymb}
\usepackage{booktabs}
\usepackage{yfonts}

\renewcommand{\theequation}
{\arabic{section}.\arabic{equation}}

\makeatletter
\def\eqnarray{ \stepcounter{equation} \let\@currentlabel=\theequation
 \global\@eqnswtrue
 \global\@eqcnt\z@
 \tabskip\@centering
 \let\\=\@eqncr
 $$\halign to \displaywidth\bgroup\@eqnsel\hskip\@centering
 $\displaystyle\tabskip\z@{##}$&\global\@eqcnt\@ne
 \hfil$\displaystyle{{}##{}}$\hfil
 &\global\@eqcnt\tw@$\displaystyle\tabskip\z@{##}$\hfil
 \tabskip\@centering&\llap{##}\tabskip\z@\cr}
\makeatother

\makeatletter
\def\@arrayacol{\edef\@preamble{\@preamble \hskip .5\arraycolsep}}
\def\array{\let\@acol\@arrayacol \let\@classz\@arrayclassz
\let\@classiv\@arrayclassiv \let\\\@arraycr\def\@halignto{}\@tabarray}
\makeatother
 
\newcommand{\be}{\begin{equation}}
\newcommand{\ee}{\end{equation}}

\newcommand{\bea}{\begin{eqnarray}}
\newcommand{\eea}{\end{eqnarray}}
\newcommand{\nn}{\nonumber}

\def\CD {{\cal D}}

\def\CH {{\cal H}}

\def\CK {{\cal K}}

\def\CM {{\cal M}}

\def\CS {{\cal S}}
\def\CT {{\cal T}}

\def\CV {{\cal V}}

\def\CZ {{\cal Z}}

\begin{document}

\setlength{\baselineskip}{7mm}

\begin{flushright}

{\tt NRCPS-HE-52-2021} \\

\end{flushright}

\vspace{1cm}
\begin{center}
{\Large ~\\{\it  Maximally Chaotic Dynamical System 

\vspace{0.3cm}
of \\
\vspace{0.3cm}
Infinite Dimensionality

}

}

\vspace{2cm}

{\sl   George Savvidy

\bigskip
  {\sl Institute of Nuclear and Particle Physics\\
  Demokritos National Research Center\\
 Ag. Paraskevi,  Athens, Greece}

\bigskip

} 
\end{center}
\vspace{1cm}

\date{\today}

\begin{abstract}
We analyse the infinite-dimensional limit of the maximally chaotic dynamical systems that are defined on N-dimensional tori. These hyperbolic systems found successful application in computer algorithms that generate high-quality pseudorandom numbers for advanced Monte Carlo simulations. The chaotic properties of these systems are increasing with $N$  because  the corresponding Kolmogorov-Sinai entropy grows  linearly with  $N$.  We calculated the spectrum and the entropy of the system that appears in the infinite dimensional limit.   We demonstrated that the limiting system has exponentially expanding and contracting foliations and therefore belongs to the Anosov C-systems of infinite dimensionality.  The liming system defines the hyperbolic evolution of the continuous functions very similar to the evolution of a velocity function describing the hydrodynamic flow of fluids. We compare the chaotic properties of the limiting system with those of the hydrodynamic flow of incompressible ideal fluid on a torus investigated by Arnold. This maximally chaotic system can find application in Monte Carlo method, statistical physics  and digital signal processing.

\vspace{1cm}
\end{abstract}

\pagestyle{plain}

\section{\it Introduction }The maximally chaotic dynamical systems continue to attract great attention of the researchers due to their rich physical properties and extended areas of application. In recent years the classical and quantum-mechanical concepts of maximally chaotic systems were developed in series of publications with application to the physics of black holes\cite{Shenker:2013pqa,Maldacena:2015waa,Gur-Ari:2015rcq,Cotler:2016fpe,Arefeva:1999, Sekino:2008he}, of non-Abelian gauge fields \cite{Savvidy:1982jk,Savvidy:2020mco} and string theory \cite{Gross:2021gsj}, to the fluid  dynamics \cite{turbul,Arakelian:1989,Endlich:2010hf} and astrophysics \cite{Chandrasekhar, body, garry, Belinsky:1982pk}, foundation of statistical physics \cite{Gibbs,hopf1,krilov,arnoldavez,kornfeld} and computer science \cite{yer1986a,konstantin,Savvidy:2015jva}.  There is a growing evidence that intrinsic properties of the Hawking black hole radiation can be understood in terms of classical and quantum theory of chaos  \cite{Hawking:1976ra,Hawking:2015qqa,Page:1993wv,Susskind:1993if,Hawking:2016msc,Strominger:2014pwa,Penington:2019kki}. It was conjectured that the resolution of the black hole information paradox of black holes radiation, which behaves as a black body radiation with finite temperature and is similar to the thermodynamic system characterised by entropy and other thermodynamic quantities, can be formulated in terms of  maximally chaotic dynamical systems \cite{Shenker:2013pqa,Maldacena:2015waa,Gur-Ari:2015rcq,Cotler:2016fpe,Arefeva:1999, Sekino:2008he,Savvidy:2020mco}. We are interested to analyse deterministic dynamical systems, which show up physical properties intrinsically related to the thermodynamical behaviour \cite{Gibbs, hopf1, krilov, arnoldavez, kornfeld, anosov1, leonov, Gutzwiller}.

There is an intuitive understanding about how strong chaotic behaviour of a classical  system can be,  but it seems natural to define a maximally chaotic system as a system that has a nonzero Kolmogorov-Sinai entropy and therefore belongs to the so called K-systems \cite{kolmo,kolmo1,sinai3,anosov,Savvidy:2020mco}.  The examples of maximally chaotic systems were investigated and constructed in the earlier investigations  of Artin, Hadamard, Hedlund, Hopf,  Birkhoff, von Neumann  and many others researchers  working in ergodic theory \cite{Artin,Hadamard, hedlund,  Hopf, rokhlin1, rokhlin2}.  A large class of maximally chaotic dynamical systems was constructed by Anosov \cite{anosov,smale,sinai2,margulis,bowen0,bowen, gines}. These are the systems that fulfil the C-condition that  leads to the exponential instability of the trajectories \cite{theproperty},   to the mixing of all orders and  to a positive Kolmogorov-Sinai  entropy.  As a result it is natural to call them C-K systems \cite{Savvidy:2020mco}.

The geodesic flows on closed Riemannian manifolds of variable negative sectional curvatures  fulfil the C-condition and therefore represent a rich class of maximally chaotic dynamical systems \cite{anosov,anosov1}.  A progress in understanding of the chaotic behaviour of the non-Abelian gauge fields, of the N-body systems in Newtonian gravity and  of some cosmological models in general relativity were achieved by the application of the methods and results of the ergodic theory and the theory of geodesic C-K flows \cite{Savvidy:1982jk, Savvidy:2020mco, Baseyan, Natalia,Chirikov,Asatrian:1982ga, SavvidyKsystem, Savvidy:1984gi,Chandrasekhar, body, garry, Belinsky:1982pk}.  

A class of hyperbolic C-K systems defined on a torus found application in computer algorithms that generate high-quality pseudorandom numbers for the advanced Monte Carlo simulations \cite{yer1986a,konstantin,Savvidy:2015jva}.   The chaotic properties of  C-K systems on a torus are increasing with the dimensionality $N$  of the systems because the entropy increases linearly with $N$  as $h(A)=  {2\over \pi}~ N$.  It seems important to investigate the infinite-dimensional limit of these systems that found successful application in Monte Carlo method \cite{hepforge,boost,clhep,root,cern} and to understand the exceptional  properties of the limiting system.

\begin{figure}
\begin{center}
\includegraphics[width=3.9cm]{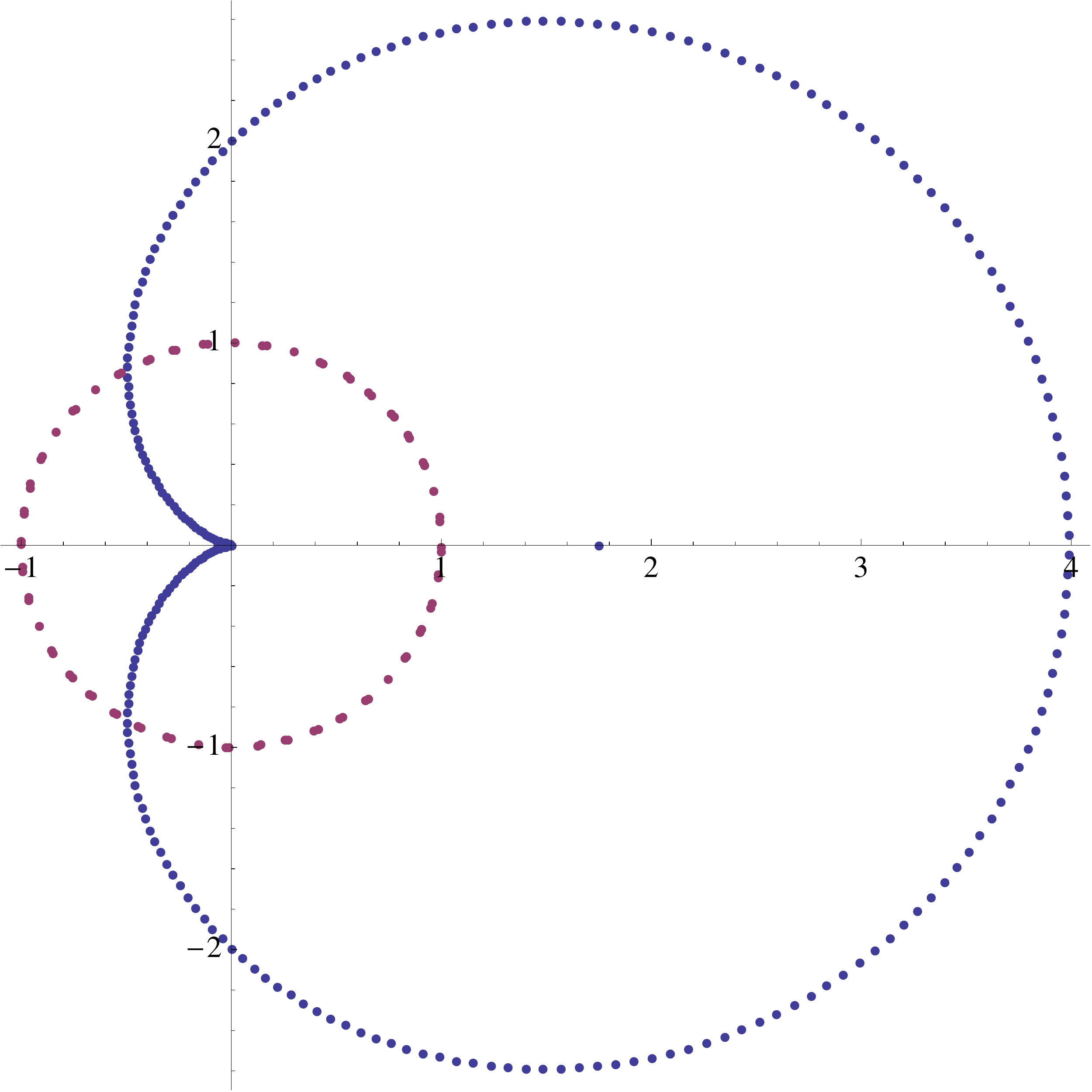}~~~~~~~~~~~~~~~~~
\includegraphics[width=4cm]{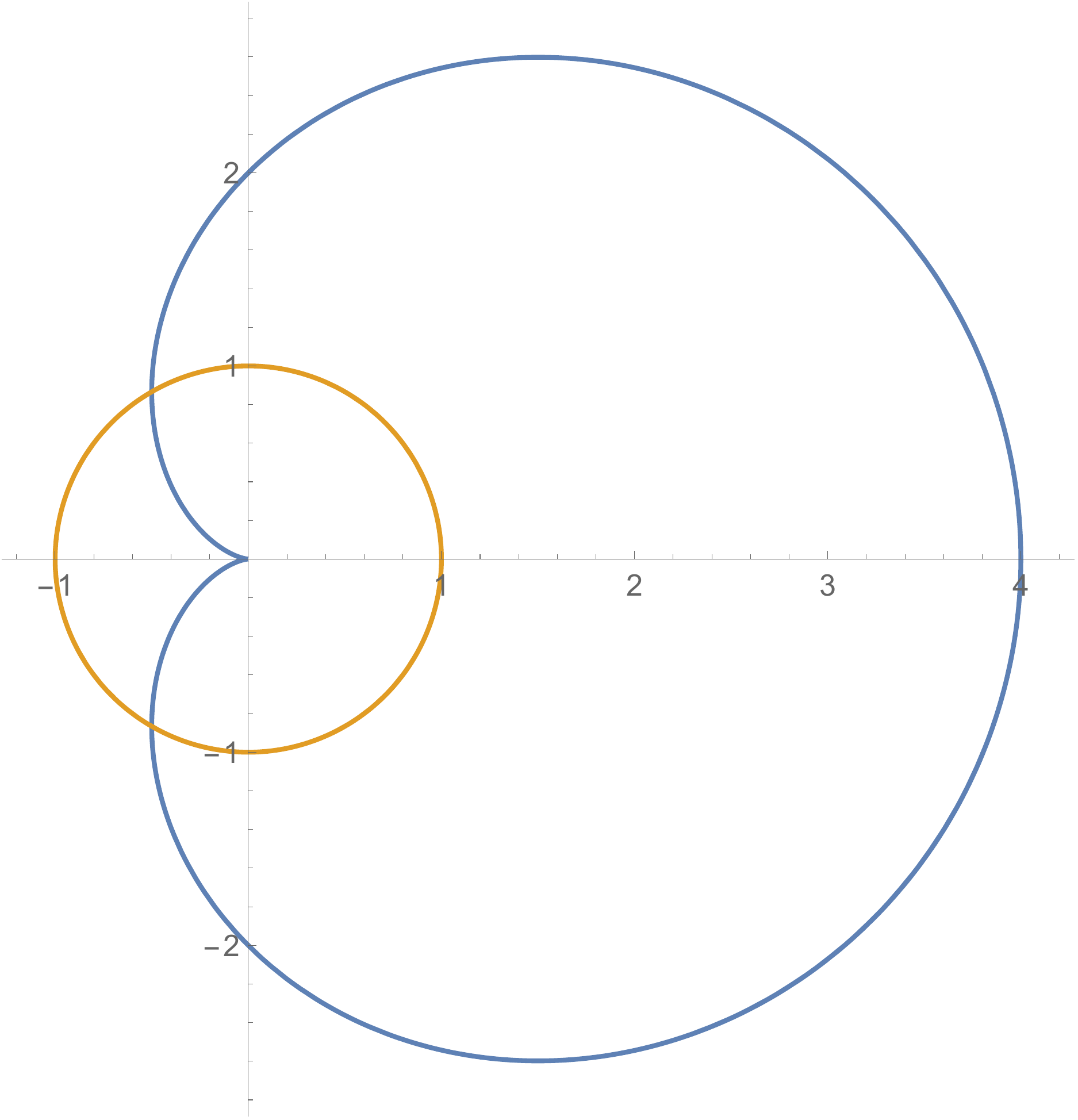}
\caption{
The left figure shows the distribution of the eigenvalues of the operators $A(N)$, where 
$N=256$. On the right figure is the distribution of the eigenvalues of the infinite-dimensional operator $A$ in  $\lambda $ plane. The  unit circles are shown in both figures.}
\label{fig1}
\end{center}
\end{figure}

\section{\it Chaotic Dynamical Systems on N-dimensional Tori}

The C-K system  that realises automorphisms of a torus  with coordinates $ (u_1,...,u_N)$ is defined as an integer matrix transformation \cite{anosov,yer1986a,konstantin,Savvidy:2015jva,sinai3,rokhlin1,rokhlin2,gines}:
\be
\label{eq:rec}
u_i^{(k+1)} = \sum_{j=1}^N A_{ij}(N) \, u_j^{(k)} ~~~~~\textrm{mod}~ 1,~~~~~~~~~k=0,1,2,...
\ee
where the components of the vector $u^{(k)}$ are   
$
u^{(k)}= (u^{(k)}_1,...,u^{(k)}_N).
$ 
The phase space of the  system is the $N$-dimensional torus $\CT^N$ 
appearing in factorisation of the Euclidean space $E^N$ with coordinates $u= (u_1,...,u_N)$ over an integer lattice $\CZ^N$ endowed with the invariant Liouville's measure  $d\mu = du_1...du_N$ \cite{anosov}.  The automorphisms (\ref{eq:rec})  fulfils  the C-condition if and only if  the integer matrix $A(N)$ has no eigenvalues on the unit circle and has the determinant equal to one, that is, the  eigenvalues $\{ \Lambda = {\lambda_1},...,
\lambda_N \}$ fulfil the following conditions:
\bea\label{mmatrix}
1)~Det  A(N)=  {\lambda_1}\,{\lambda_2}...{\lambda_N}=1,~~~~~
2)~~\vert {\lambda_i} \vert \neq 1, ~~~\forall ~~i .~~~~~~
\eea
The eigenvalues  are divided into sets  with modulus smaller and larger than one:
\bea\label{eigenvalues1}
0 <  \vert \lambda_{\alpha} \vert   < 1   & \textrm{ for } \alpha=1...d,~~~~~~~~~
1 <  \vert \lambda_{\beta} \vert  < \infty & \textrm{ for } \beta=d{+}1...N . 
\eea
The C-K system (\ref{eq:rec})   has nonzero entropy and  its value  $h(A)$ can be calculated in
terms of its  eigenvalues \cite{anosov,yer1986a}:
\be\label{entropyofA}
h(A) = \sum_{\beta   } \ln \vert \lambda_{\beta} \vert =-\sum_{\alpha   } \ln { \vert \lambda_{\alpha} \vert}.
\ee
The eigenvalues $\lambda_{\beta}$ larger than one are nothing else but the Lyapunov exponents and characterise the chaotic properties of the system as it follows from the above definition of entropy (\ref{entropyofA}). 

We are interested to consider the infinite-dimensional limit $N\rightarrow \infty $ of the system  (\ref{eq:rec}) when the operator $A(N)$ is given by the $N\times N$ matrix with all integer entries  
 $A_{ij} \in \mathbb{Z}$ and has the following form \cite{yer1986a,konstantin,Savvidy:2015jva}:
\be
\label{eq:matrix}
A(N) =
   \begin{pmatrix}
      1 & 1 & 1 & 1 & ... &1& 1 \\
      1 & 2 & 1 & 1 & ... &1& 1 \\
      &&&...&&&\\
      1 & N & N{-}1 &  ~N{-}2 & ... & 3 & 2
   \end{pmatrix}.
\ee
It has the determinant equal to one and the spectrum of inverse matrix has the form \cite{Savvidy:2015jva}
\bea\label{eq:curve}
&\lambda_j = 1 -2 \exp(i\, \pi j/N) + \exp(2i\, \pi j/N),~~~ 
& ~ j=-N, -N+2,...N-2, N ~
\eea
shown  in Fig.\ref{fig1}. The kernel  $A(N) u_0=0$ of the operator $A(N)$ consists of only one vector with zero components $u_0= (0,...,0)$. The eigenvalues  fulfil the C-condition (\ref{mmatrix})
and the entropy of the system  can be calculated for the
large values of $N$  as a sum over eigenvalues:
\be\label{linear}
h(A)= \sum_{\beta   } \ln \vert  \lambda_{\beta} \vert = \sum_{-2\pi/3 <  \phi_j < 2\pi/3} \ln (4\cos^2(\phi_j/2)~ \sim
{2\over \pi}~ N,
\ee
where the eigenvalues are given in (\ref{eq:curve}) and  $\phi_j = \pi {j\over N} $. The entropy increases linearly with the dimension $N$ of the operator $A(N)$. We note that the special form of the matrix $A(N)$ in \eqref{eq:matrix} has highly desirable property of having a widely spread, nearly continuum spectrum of eigenvalues (\ref{eq:curve}) shown in Fig.\ref{fig1} and indicating that the exponential mixing takes place in many scales \cite{yer1986a}.

\section{\it Infinite-dimensional Limit of $A(N)$ System}
We are interested to define and investigate  the system  that appears in the infinite-dimensional limit of $A(N)$ when $N \rightarrow \infty$. The size of state vector $u= (u_1,...,u_N)$ tends to infinity and it seems natural to expect that it can be represented by a continuous function $\psi$ defined in an infinite-dimensional space $\CH=\{\psi(u)\}$, possibly a Hilbert space, and the operator $A(N)$ will reduce to a differential operator $A$ acting in $\CH$. The existence of such a limit would mean that $A$ defines a "measure preserving"  transformation of continuous functions $  A^n \psi = \psi^{(n)} $ in $\CH$ that will have maximally strong chaotic properties. Our intension is to define this limiting system, to explore its properties and possible applications in Monte Carlo method.  It seems that the investigation of infinite-dimensional "fully chaotic" transformation of continuous functions may also help to understand better a chaotic/turbulent motion of fluids.  As it was demonstrated by Arnold \cite{turbul,Arakelian:1989},  the solutions of the partial differential equation describing the evolution  of the hydrodynamic flow of incompressible ideal fluid can be considered as a continuous measure preserving transformation of fluid velocity and the evolution is partially chaotic, that is, the flow is exponentially unstable in some directions and is stable in other directions (the details will be discussed in the last paragraph).

In many areas of mathematics  the consideration of infinite-dimensional limits is an ambiguous procedure, and a priory there is no guarantee that a sensible limit of a given finite-dimensional structure exists. In our case the dimension of the N-dimensional torus $S^1 \otimes....\otimes S^1$  tends to infinity and it is unclear what type of phase space should be taken in the limit.  The hint that a sensible limit may exist comes from the fact that as $N \rightarrow \infty$ the eigenvalues (\ref{eq:curve}) fill out the cardioid curve more and more dense without producing any deformation of the cardioid curve that can be seen in figure Fig.\ref{fig1} \cite{Savvidy:2015jva}. The other important hint is that the inverse matrix $A(N)^{-1}$ is reminiscent to the matrix that represents the discrete version of the second-order differential operator \cite{Savvidy:2015jva}. If one supposes that the state vector becomes a function $\psi(x)$ with its argument on a real line $x \in R^1$, then it seems natural to look for a differential operator of the second order and the one that will reproduce the eigenvalue spectrum distributed on cardioid  curve.  Having in mind the above consideration let us consider the differential operator
\be\label{limitings}
A_x= 1 - 2 \exp{\Big({d \over d x}\Big)} + \exp{ \Big(2 {d \over d x}\Big)}= {d^2 \over d x^2} +{d^3 \over d x^3} +{7\over 12}{d^4 \over d x^4} +...
\ee
 acting in the Hilbert space of functions $\CH=\{\psi(x)\}$ defined on the interval $x \in [-\infty, +\infty]$. The series expansion of the operator has indeed a second-order differential operator and also infinitely many high derivative terms. Its spectral characteristics are defined by the eigenvalue equation 
$ 
A_x \psi(x) = \lambda \psi(x).
$
Searching the eigenfunctions in the form of plane waves 
\be\label{eigenfun}
\psi_a(x)= e^{i a x}
\ee
one can find that the spectrum represents a continuous cardioid curve on the complex plane
\be\label{eigenvalues12}
\lambda(a) = 1-2 e^{i a} + e^{2 i a}= 4 e^{i (a+\pi)} \sin^2{\Big({a\over 2}\Big)}.
\ee
It is similar to the discrete spectrum (\ref{eq:curve}) and has the periodic structure 
\be\label{period}
\lambda(a+2\pi k) =\lambda(a), ~~k=0,\pm 1, \pm2, ....
\ee
The  spectrum is continuous and the eigenvalues are distributed in the complex plane representing the cardioid  curve shown in Fig.\ref{fig1}. As the real momentum parameter $a$ varies on the real line interval  the eigenvalues run around the cardioid infinitely many time (\ref{period}). The eigenvalues of the operator $A_x$ can be divided into two sets   $\{ \lambda_{\alpha}  \} $ and $\{  \lambda_{\beta }  \} $
with modulus smaller and larger than one:
$
0 <  \vert \lambda_{\alpha} \vert   < 1,~~
1 <  \vert \lambda_{\beta} \vert   . 
$
The eigenvalues $\lambda_{\alpha}$ and  $\lambda_{\beta}$ can be found  using (\ref{eigenvalues12}):
\bea\label{lessone}
\lambda_{\alpha} &=& 4 e^{i (a+\pi)} \sin^2{({a\over 2})} ,~~\text{when}~~ 
- {\pi \over 3} +2\pi k   < a < + {\pi \over 3} +2\pi k , ~~~~~\nn\\
\label{largeone}
\lambda_{\beta} &=& 4 e^{i (a+\pi)} \sin^2{({a\over 2})} ,~~\text{when}~~   + {\pi \over 3} +2\pi k   < a < + {5 \pi \over 3} +2\pi k,
\eea
where $k=0,\pm 1, \pm2, ....$.This structure of the spectrum repeats itself with the period $2\pi$.
There are  eigenvalues $\vert \lambda \vert =1$ corresponding to  $a=\pm {\pi \over 3} +2\pi k$ where the cardioid intersects a unit circle shown  in Fig.\ref{fig1}.

\section{\it The Kolmogorov-Sinai Entropy of Limiting System}
In order to establish the fact that the operator $A_x$ is defining a measure-preserving transformation one should calculate the determinant of the operator $A_x$. The measure-preserving transformations of the phase spaces is a characteristic property of Hamiltonian systems that is expressed in terms of the Liouville's theorem and represent a large class of dynamical systems that are considered in ergodic theory \cite{Gibbs,hopf1,krilov,arnoldavez,kornfeld}.  Using the fact that $\ln Det A_x = Tr \ln A_x $ we will have 
\be\label{det}
\ln Det A_x= \sum^{\infty}_{k=-\infty} \int^{+\pi +2\pi k }_{-\pi +2\pi k} \ln[  e^{i (a+\pi)}  4 \sin^2{\Big({a\over 2}\Big)} ]{d a \over 2 \pi}=0,
\ee 
that is, the determinant is equal to one $Det A_x=1$ and the operator $A_x$ is defining  a measure-preserving transformation (the kernel subspace $\CK$ of the operator  $A_x$  will be defined  in (\ref{kernel})). One can define now the homomorphism of the Hilbert space  $\CH=\{\psi(x)\}$ in terms of the operator $A_x$  as 
\be\label{direct}
\psi^{(1)}(x) = A_x \psi(x)
\ee
and the dynamical system on the infinite dimensional phase space $\CH$ as: 
\be\label{auth}
\psi^{(n)}(x) = A_x^n  \psi(x)~~~~~\textrm{mod}~ 1 ,~~~~~~~~n=0,1,2,....
\ee
 The homomorphism  (\ref{auth}) is defined by mod 1 operation meaning that the functions $\psi^{(n)}(x)$ are  wrapping around an infinitely long cylinder\cite{onecan} $R^1 \otimes S^1$   ($\psi: R^1 \rightarrow S^1$).

The determinant of the operator $A$ is equal to one and the eigenvalues are distributed inside and outside of the unit circle, and we have an example of infinite-dimensional hyperbolic C-K system of the type (\ref{mmatrix}). To get convinced that the operator $A_x$ represents a hyperbolic C-K system one should establish the existence of exponentially expanding and contracting foliations \cite{anosov}. Let us consider  the evolution of the infinitesimal perturbation $\psi \rightarrow \psi + \delta \psi $ under the  action of $A$ operator in analogy with the geodesic deviation equation: 
\be\label{primarysys}
\delta\psi^{(n)}(x) = A_x^n  \delta\psi(x).
\ee
The  deviation $\delta L$ can be evaluated by using the standard inner product in Hilbert space and the mean value theorem \cite{if}:
\bea
\delta L_n &=&   \langle  \delta\psi^+\vert A_x^n  \delta\psi \rangle  =\int^{+\infty}_{-\infty} \delta\psi^+(x) A_x^n  \delta\psi(x) dx  \nn\\
&=& \int^{+\infty}_{-\infty} \int_{\Delta a}
 {da^{'} \over 2 \pi}\int_{\Delta a} {da \over 2 \pi}~ e^{- i a^{'} x} \delta\phi^+(a^{'}) ~  \Big[4  \sin^2{({a\over 2})} \Big]^n ~ e^{in (a+\pi)} e^{ i a x }   \delta\phi(a)~  
dx \nn\\
&=&  {1\over 2\pi }\int_{\Delta a} da  ~    e^{in (a+\pi)} ~ \Big[4  \sin^2{({a\over 2})} \Big]^n   ~\vert \delta\phi(a) \vert^2   ~ =  ~e^{in (\bar{a}+\pi)}  e^{ n \ln{[4  \sin^2{({\bar{a}\over 2})}]} }~  \overline{\vert \delta\phi^2 \vert}, 
\eea 
where $\bar{a}$ is a  number in the interval $\Delta a \subset ( \pi/3,5\pi/3 ) $ and 
$ \overline{\vert \delta\phi^2 \vert}  =   {1 \over 2 \pi} \int_{\Delta a}  \vert \delta\phi \vert^2  da $.
The absolute value of the deviation  is growing exponentially with the iteration time  $n$  as 
\be
\vert \delta L_n \vert \sim  \overline{\vert \delta\phi^2 \vert} ~e^{ n \ln{[4  \sin^2{({\bar{a}\over 2})}]} } .
\ee 
In the above perturbation the Fourier  spectrum of the  function  $\delta\psi(x)$ is localised in the region of the spectrum $  \Delta a $,   where the eigenvalues $\lambda_{\beta}$ are larger than one, $\Delta a \subset ( \pi/3,5\pi/3 )~ $,  and are defined in (\ref{largeone}).  The integration over $a$ and $a^{'}$ was specified to be in that region $  \Delta a $  and the perturbation function had the following form:
\be
 \delta\psi(x) =   \int_{ \Delta a}  \delta\phi(a)    e^{i a x}     {da\over 2 \pi}.
\ee
In a similar way one can get convinced that the exponential contraction takes place when the Fourier spectrum of the perturbation function is localised in the part of the spectrum $\lambda_{\alpha}$, where the eigenvalues are less than one $\Delta a \subset ( -\pi/3,\pi/3 )$ in (\ref{lessone}). The existence of exponentially expanding and contracting foliations is a sufficient condition for a dynamical system to expose strong statistical/chaotic  properties and, in particular, to have nonzero Kolmogorov-Sinai entropy  and to be classified as a  hyperbolic C-K system. In physical terms this means that the Fourier amplitudes $ \phi(a)$ of the initial state vector $\psi(x)$ are stretched and compressed depending on whether the value of $a$  is in the interval $a \in ( {\pi \over 3} +2\pi k ,  {5 \pi \over 3}+2\pi k) $ or in the interval   $a \in ( -{\pi \over 3}+2\pi k,{\pi \over 3} +2\pi k) $, where $~k=0,\pm 1, \pm2, ....$.

\begin{figure}
 \begin{center}
\includegraphics[width=5cm]{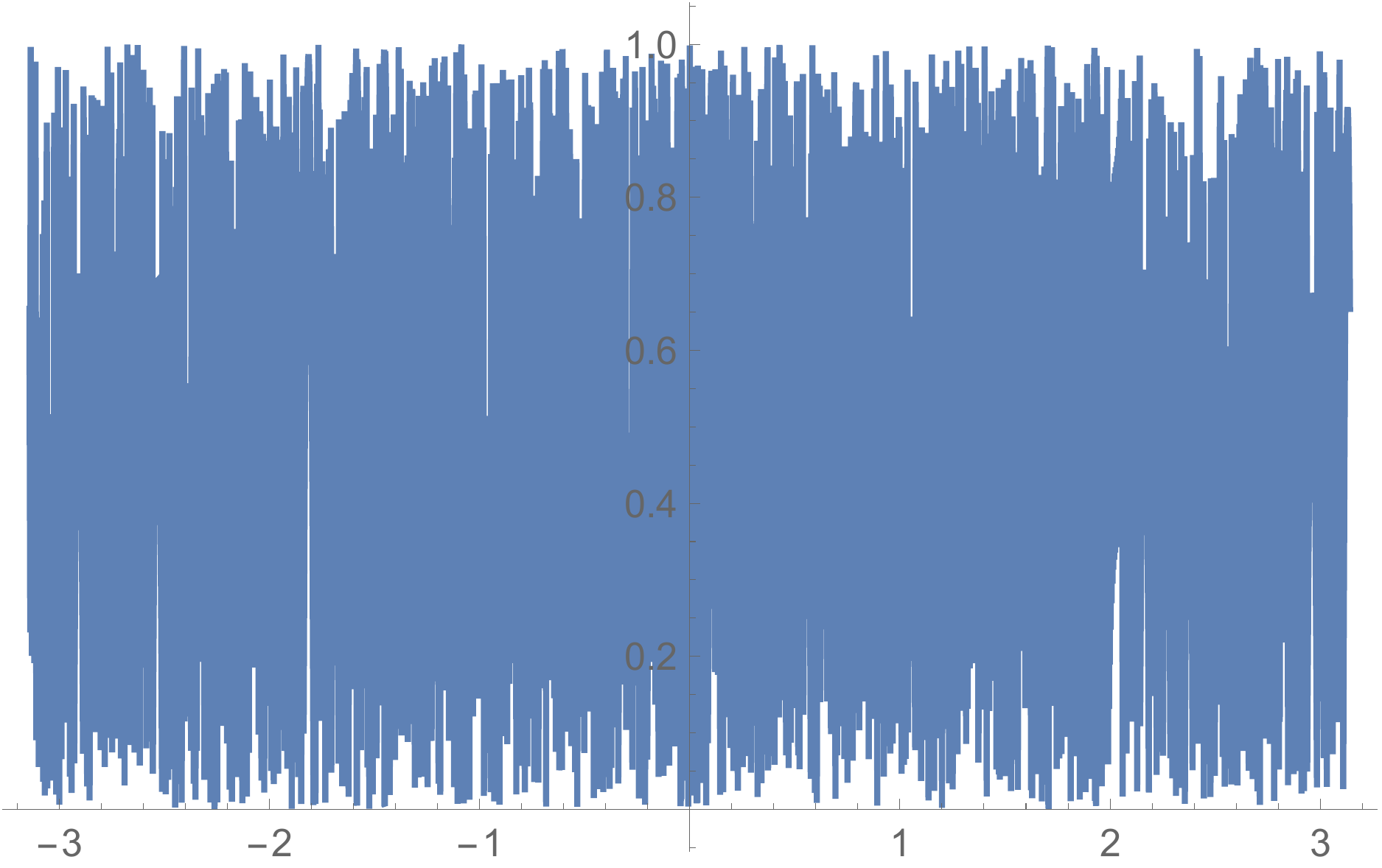}
\end{center}
\caption{\label{fig3}
The figure demonstrates the result of a triple iteration (\ref{auth}) $\psi^{(3)}(x)= A_x^3 \psi(x)$ of the smooth function $\psi(x) = \sin(x) +\sin(2x) - \cos(3x) +\sin(4x)$.}
\end{figure}

The above consideration allows to  calculate the Kolmogorov entropy of the system per unit of the iteration time $n$ as it is was defined by Kolmogorov \cite{kolmo1}.  The new aspect that appears  in this infinite-dimensional system case is that the spectrum (\ref{eigenvalues12}) is continuous and repeats itself infinitely many times. For that reason the standard Kolmogorov definition \cite{kolmo1} of the entropy per unit iteration time is equal to infinity.  That can be observed also from the equation (\ref{linear}) when $N \rightarrow \infty$. In this circumstances  one can propose to calculate the entropy per unit period  (\ref{period}) of the spectrum  (\ref{eigenvalues12}):
\be\label{entropyofAA}
h(A_x)=\sum_{\alpha   } \ln {1\over \vert \lambda_{\alpha} \vert}=   -\int^{\pi/3}_{-\pi/3} \ln [4\sin^2({a\over 2})]{da \over 2\pi}=
2 i  [Li_2(e^{i5\pi/3}) -  Li_2(e^{i\pi/3})] \sim {2\over \pi} ,
\ee
 where we used the fact that   $ \prod_{\alpha} \lambda_{\alpha}   \prod_{\beta} \lambda_{\beta} =1$ and $Li_n(z)$ is the polylogarithm function. This result is understandable in the sense that the finite-dimensional matrix  system (\ref{eq:matrix}) considered above  had the entropy $\sim {2\over \pi} N$,  where $N$ is the dimension of the matrix operator.  As far as the operator $A$  can be considered  as the infinite dimensional limit $N\rightarrow \infty$ of (\ref{eq:matrix}), the standard Kolmogorov entropy of the system (\ref{auth}) tends to infinity but its entropy "per spectral period" is finite.  The Fig.(\ref{fig3}) demonstrates an example of the iteration (\ref{auth}) of a smooth function. 

The inverse operator $G$ is defined by the  equation 
\be\label{inverse}
A_x ~G(x-y) =  \sum^{\infty}_{k=-\infty} \delta(x-y-k)
\ee
and has the following solution: 
\be\label{fixed}
G(x-y) =  {1\over 2} (x-y)^2 \sum^{\infty}_{k=-\infty} e^{2 \pi i k (x-y)}.
\ee
The general solution of (\ref{inverse}) can be expressed as a sum the fixed solution (\ref{fixed}) and an arbitrary element of the kernel $\CK$.  The kernel subspace $\CK$ of the operator $A_x$ is defined by the equation $A \psi_0(x)=0$ and has the following solution:
\be\label{kernel}
 \psi_0(x) = (c_1 x +c_2) \sum^{\infty}_{k=-\infty} a_k e^{2 \pi k i x},
\ee
where $c_1,c_2, a_k$ are arbitrary constants. The spectrum belongs to the discrete values of spectral parameter $a=2\pi k$ in (\ref{eigenfun}) and have a zero measure in the continuous spectrum of the limiting system (\ref{limitings}). The inverse transformation $\psi^{(-n)}   = G^n \psi,~ n=0,1,2,...$ is therefore defined modulo kernel (\ref{kernel}).  We will defined it in its most simple form (\ref{fixed}):
\be\label{inversetime}
\psi^{(-1)}  = \int G(x-y) \psi(y) dy = \sum^{\infty}_{k=-\infty} {1\over 2} k^2 ~\psi(x-k)~~~~~\textrm{mod}~ 1.
\ee
Thus the evolution of the system is given in both "time directions" by (\ref{inversetime})  and  (\ref{direct}).

Because the determinant of the operator $A_x$ is equal to one on a quotient space $\CH/\CK$ \cite{thereare} of functions  on a cylinder $R^1 \otimes S^1 $, where $\CK$ is a kernel (\ref{kernel}),  it is natural to think that the operator  $A_x$ defines a measure preserving  transformation.  In this circumstances one can try to define a measure  that is  invariant with respect to the transformations generated by $A_x$.  The volume element $ \CV$ in the quotient space $\CH/\CK$ can be defined by using the Wiener-Feynman functional integral
$
 \CV_C= \int_{C} F[\psi] \CD \psi(x),
$ 
where $C $ is a subset in $\CH/\CK$ and the functional $F[A_x \psi]=F[\psi]$  should be  invariant under the action of the transformations generated by the operator $A_x$. The measure $\CD \psi(x)$ is invariant because the determinant of the corresponding Jacobian operator is equal  to one (\ref{det}). The invariant functional $F[\psi]$ can be constructed by projecting the state vectors  on the eigenfunctions   $\phi( \pm{\pi \over 3} +2\pi k )=\int^{\infty}_{-\infty} e^{ i ( \pm {\pi \over 3} +2\pi k)  x } \psi(x) dx$ (\ref{eigenfun}) that correspond to the eigenvalues  of modulus one $\vert \lambda(\pm{\pi \over 3} + 2 \pi k)\vert  =1$, $k=0,\pm 1, \pm2, ....$.  The invariant functional $F[\psi]$ can be defined as:  
\be
F[\psi] = \exp{  \Big[-  \sum_{k =\pm1 ,\pm2,...}{  \vert    \phi({\pi \over 3} + 2 \pi k) \vert^2    +\vert    \phi(-{\pi \over 3} + 2 \pi k) \vert^2   
   \over 2}   ~ \Big] }.
\ee

\section{\it Flow of Incompressible Ideal Fluid}
Were similar systems investigated and successfully used  in the past? The solutions of the partial differential equation describing the evolution $t \rightarrow g_t(x)$ of the hydrodynamic flow of incompressible ideal fluid filled in a two dimensional torus $x \in \CT^2$ can be considered as a continuous area preserving diffeomorphims $SDiff(\CT^2)$ of a torus $\CT^2 \rightarrow \CT^2$.  In Arnold approach \cite{turbul} the ideal fluid flow is described by the geodesics $g_t(x) \in G$ on the  diffeomerphism group $G=SDiff(\CT^2)$ with the velocity $v_t(x)= \dot{g}_t(x) g^{-1}_t(x)$ belonging  to the corresponding algebra \textfrak{g}=sdiff$(\CT^2)$  of divergence free vector fields. The Riemannian metric on the group $G$ is induced from the metric on a torus \cite{turbul} and the stability of the geodesic flows on  the group $G$  can be analysed by investigating the behaviour of the corresponding sectional curvatures $K(v,\delta v)$ \cite{turbul,anosov,app}.  It was found that the  flows that are defined by a parallel velocity field on $\CT^2$ are unstable because the sectional curvatures are negative and the flow is exponentially unstable.  In other directions the sectional curvatures are positive and the flows are stable  \cite{turbul}.   A similar  stability analysis was performed  for the hydrodynamic flow on two-dimensional sphere  $\CS^2$ in \cite{Arakelian:1989,Smolentsev} and on high-dimensional torus $\CT^N$ in  \cite{Lukatzki:1981,Smolentsev}. In all these  cases  the flow is exponentially unstable in some directions and is stable in some other directions, resulting in the limitation of  predictability of the hydrodynamic flow and  leading to the principal difficulties of a long-term "weather forecasting" \cite{turbul}.  Comparing these systems with the system considered above one can observe that here  we have discrete in time transformations of the phase space and, secondly, the system (\ref{limitings}),  (\ref{direct}), (\ref{inversetime}) shows up  exponential instability of its geodesics in the full quotient phase space $\CH/\CK$. This full phase space chaotic behaviour can find application in Monte Carlo method, statistical physics  and in digital signal processing  in communication systems,  the subjects of future investigation. 

\section{\it Conclusion}
We were able to define the infinite-dimensional limit  of maximally chaotic dynamical systems on N-dimensional tori when the dimension $N$ tends to infinity.  The limiting system $A_x$ is represented by a nonlocal differential operator acting on the infinite-dimensional Hilbert space of continuous functions. We calculated the eigenvalue spectrum of the limiting operator, as well as  its determinant and the corresponding Kolmogorov-Sinai entropy. We investigated the exponentially expanding and contracting foliations and demonstrated that the limiting system belongs to the class of hyperbolic Anosov C-systems.  The limiting system represents a unique example of a C-system of infinite dimensionality, which has a quite simple form and is chaotic in its full phase space.

\section{\it Acknowledgement } The author would like to thank Konstantin Savvidy for stimulating discussions. This work was supported by the Horizon 2020
research and innovation programme under the Marie Sk\'lodowska-Curie
Grant Agreement No 644121.  

\section{\it Appendix}
The motion of an abstract "rigid body" rotating in high-dimensional Euclidean  space $\vec{r} \in E$ that is invariant under the isometry group $g(\xi) \in G$ can be described in terms of geodesic flow on a corresponding group manifold  \cite{turbul}.  The stationary frame coordinates  of the "rigid body" are defined as $\vec{r}= g_t \vec{Q}$,  where $g_t = g(\xi(t))$ is a time-dependent  element of the matrix group $G$ and $ \vec{Q} $ are the frame coordinates rigidly fixed to the rotating "body"  $\dot{\vec{Q}}=0$. Thus 
 \be\label{velocity}
 \dot{\vec{r}} = \dot{g}_t \vec{Q}=  \dot{g}_t   g^{-1}_t ~ g_t\vec{Q} =  \hat{\omega}_s \vec{r}.
 \ee
The matrix of angular velocity in stationary frame
 \be\label{rangvel}
 \hat{\omega}_s= \dot{g}_t g^{-1}_t 
 \ee
is a right-invariant one-form ( $d ( g g_0)( g g_0 )^{-1}   =  d  g g^{-1}$, where $g_0$ is a fixed element of the group $G$). The matrix of angular velocity in rotating frame  
 \be\label{langvel}
 \hat{\Omega}_c= g^{-1}_t \dot{g}_t 
 \ee
is left invariant  because $(g_0 g)^{-1} d (g_0 g)   =  g^{-1} d  g  $. It follows that\footnote{The general relation between operators in stationary and rotating frames is $A_s = g_t  A_c g^{-1}_t $ . }
\be\label{angtrans}
 \hat{\omega}_s= \dot{g}_t g^{-1}_t = g_t  g^{-1}_t \dot{g}_t g^{-1}_t  = g_t  \hat{\Omega}_c g^{-1}_t .
 \ee
The kinetic energy is defined as a sum of the kinetic energies of all "parts" of the rotating "body" through the velocities  (\ref{velocity}):  
\be\label{kinener}
 T= {1\over 2}\ \sum_a m_a \dot{\vec{r}}_a \dot{\vec{r}}_a   =  {1\over 2}\ Tr (  \hat{I}_s  \hat{\omega}_s ~\hat{\omega}^+_s)  = -{1\over 2}\ Tr (  \hat{I}_s  \hat{\omega}_s~ \hat{\omega}_s),
\ee 
 where one should use the relation  $\vec{r}= g_t \vec{Q}$, and therefore
 \be\label{inertiatens}
 \hat{I}_s = \sum_a m_a r_a  r^+_a = g_t  \hat{I} g^{-1}_t  , ~~~~\hat{I}  = \sum_a m_a Q_a  Q^+_a.
 \ee
 The matrix  $ \hat{I}$ is a symmetric positive definite constant  matrix that determines the "moment of inertia"  in the frame rigidly fixed to the rotating "body".  The matrix of angular momentum in stationary frame is
\be\label{mominet}
\hat{m}_s=    \hat{I}_s \  \hat{\omega}_s 
\ee
and the corresponding angular momentum in rotating frame can be defined by projection of $\hat{m}_s$ into the rotating frame:
\be\label{transfor}
 \hat{m}_s= g_t \hat{M}_c g^{-1}_t , 
 \ee
thus\footnote{The square of angular momentum is conserved: 
$Tr (\hat{m}^2_s ) = Tr [(g_t \hat{M}_c g^{-1}_t)^2] = Tr (\hat{M}^2_c).$}
\be
\hat{M}_c=   \hat{I} \  \hat{\Omega}_c ,
\ee 
where one should use the relations  (\ref{mominet}), (\ref{inertiatens}) and  (\ref{rangvel}), (\ref{langvel}).
In terms of rotating frame coordinates the kinetic energy (\ref{kinener})  will take the form  
\be\label{kinener1}
 T= - {1\over 2}\ Tr (  \hat{I}_s  \hat{\omega}_s ~\hat{\omega}_s)= -{1\over 2}\  Tr (  \hat{m}_s ~  \hat{\omega}_s ) = -{1\over 2}\  Tr (\hat{M}_c ~\hat{\Omega}_c) =  -{1\over 2} \ Tr (  \hat{I} \hat{\Omega}_c  ~\hat{\Omega}_c),
\ee
where we used the relations (\ref{transfor})  and (\ref{angtrans}).  As it follows from (\ref{kinener1}) and (\ref{Riemann-metric}), in mathematical terms the matrix  $\hat{I}$  defines  the alternative Euclidean structure on the group algebra  $\langle a,b\rangle_I= Tr(T_a \hat{I }T_b)$, where $T^a$ are the generators of the algebra \textfrak{g} and $\hat{\Omega}_c= \sum_aT_a \Omega^{a}$.

 Using the relation (\ref{transfor}) and the conservation of the angular momentum in stationary frame  $\dot{\hat{m}}_s=0$ one can get the generalised Euler equation  
\be
\dot{g}_t \hat{M}_c g^{-1}_t  + g_t \dot{\hat{M}}_c g^{-1}_t - g_t \hat{M}_c g^{-1}_t \dot{g}_t  g^{-1}_t =0,\nn
\ee
wgich can be represented in the following standard form: 
\be
\dot{\hat{M}}_c +    \hat{\Omega}_c  \hat{M}_c  -    \hat{M}_c  \hat{\Omega}_c =0,\nn
\ee
or equivalently as
\be
{d \hat{M}_c \over dt }= [\hat{M}_c ; \hat{\Omega}_c].
\ee 
In terms  of  angular velocity it takes the following form:
\be
  \hat{I} ~ {d \hat{\Omega}_c \over dt }= [   \hat{I} \hat{\Omega}_c ; \hat{\Omega}_c]~,~~~
~~~~
   {d \hat{\Omega}_c \over dt }=  \hat{I}^{-1}~ [   \hat{I} \hat{\Omega}_c ; \hat{\Omega}_c] = \Gamma ( \hat{\Omega}_c , \hat{\Omega}_c).
\ee 
The last Euler equation can be represented in the form of geodesic equation 
\be
 {d \Omega^{a} \over dt } + \Gamma^{a}_{bd}\  \Omega^{b} \ \Omega^{d} =0,
\ee
where $\Gamma^{a}_{bd}$  are the Christopher symbols of the metric (\ref{Riemann-metric}). The kinetic energy defines the left invariant metric on the group: 
\be
ds^2 =  Tr (  \hat{I}  \hat{\Omega}_c ~ \hat{\Omega}_c) dt^2=    Tr (  \hat{I} ~g^{-1} d  g \  g^{-1} d  g    )=  
Tr (  \hat{I} ~g^{-1}  {\partial g \over \partial \xi^a}  ~ g^{-1}  {\partial g \over \partial \xi^b}   ) ~d\xi^a d\xi^b,
\ee 
where the $\xi^a$ are parameters of the Lie group $G$ and 
\be\label{Riemann-metric}
ds^2 = g_{ab} ~d\xi^a d\xi^b,~~~~    g_{ab}= Tr (  g^{-1}  {\partial g \over \partial \xi^b}  ~ \hat{I} ~g^{-1}  {\partial g \over \partial \xi^a}  ~  ).
\ee
The calculation of the components of the Riemann tensor and of the sectional curvatures  
\be
K(\xi,\eta) = { R_{abcd} \xi^a \eta^b  \xi^c \eta^d \over  \vert \xi \wedge  \eta   \vert^2 }
\ee
allows to investigate the stability of the geodesic flows  even in the cases when $G$ is the infinite-dimensional group of diffeomorphims $SDiff(\CM)$ that describes the flow of incompressible ideal fluid filled in a   manifold $\CM$ \cite{turbul,Arakelian:1989,Lukatzki:1981,Smolentsev}.

\section{\it Data Availability Statement. } The data that support the findings of this study are available within the article and also in the references \cite{yer1986a,konstantin,Savvidy:2015jva,hepforge,boost,clhep,root,cern}.

\vfill
\bibliography{your-bib-file}

\end{document}